\documentclass[preprints,article,accept,moreauthors,pdftex,10pt,a4paper]{mdpi}

\def\be{\begin{eqnarray}}   
\def\ee{\end{eqnarray}}

\firstpage{1} 
\pubvolume{xx}
\issuenum{1}
\articlenumber{1}
\pubyear{2018}
\copyrightyear{2018}
\history{Received: date; Accepted: date; Published: date}

\usepackage{color}
\Title{\textit{Ab initio} simulation of attosecond transient absorption spectroscopy in two-dimensional materials}
\Author{Shunsuke A. Sato$^{1}$*, Hannes H\"ubener $^{1}$,
Umberto De Giovannini$^{1}$, and Angel Rubio $^{1,2}$}

\AuthorNames{Firstname Lastname, Firstname Lastname and Firstname Lastname}

\address{%
$^{1}$ \quad Max Planck Institute for the Structure and Dynamics of Matter and Center for Free-Electron Laser  Science, Luruper Chaussee 149, 22761 Hamburg, Germany\\
$^{2}$ \quad Center for Computational Quantum Physics (CCQ), The Flatiron Institute, 162 Fifth avenue, New York NY 10010}

\corres{Correspondence: shunsuke.sato@mpsd.mpg.de}

\abstract{We extend the first-principles analysis of attosecond transient absorption
spectroscopy to two-dimensional materials.
As an example of two-dimensional materials, we apply the analysis to 
monolayer hexagonal boron nitride (\textit{h}-BN) and compute
its transient optical properties
under intense few-cycle infrared laser pulses.
Nonadiabatic features are observed
in the computed transient absorption spectra.
To elucidate the microscopic origin of these features, we analyze
the electronic structure of \textit{h}-BN with density functional theory
and investigate the dynamics of specific energy bands with a simple two-band model.
Finally, we find that laser-induced intraband transitions play a significant role
in the transient absorption even for the two-dimensional material and that
the nonadiabatic features are induced by 
the dynamical Franz-Keldysh effect with an anomalous band dispersion.
}

\keyword{attosecond transient absorption spectroscopy; 
time-dependent density functional theory; 
first-principles simulation}

\begin{document}
%%%%%%%%%%%%%%%%%%%%%%%%%%%%%%%%%%%%%%%%%%
%% Only for the journal Gels: Please place the Experimental Section after the Conclusions

%%%%%%%%%%%%%%%%%%%%%%%%%%%%%%%%%%%%%%%%%%
\section{Introduction \label{sec:intro}}

Attosecond transient absorption spectroscopy (ATAS) is a technique that employs
attosecond laser pulses to probe modifications to the optical absorption in the time domain.
Because its attosecond time-resolution is
shorter than a typical time-scale of electron dynamics,
it can naturally capture ultrafast electron dynamics in matter.
Therefore, ATAS is one of the key experimental techniques to explore ultrafast phenomena
where nonlinear electron dynamics plays a significant role.
Indeed, ATAS has been extensively applied to atoms and molecules over the past decade
to investigate ultrafast electron dynamics in such relatively small systems
\cite{Goulielmakis2010,PhysRevLett.105.143002,PhysRevLett.106.123601,
beck2014attosecond,warrick2016probing,reduzzi2016observation}. 
Recently, ATAS has been further extended to solid-state materials
to investigate rather complex electron dynamics in various phenomena such as
band-gap renormalization \cite{Schultze1348}, 
petahertz optical drive \cite{Mashiko2016},
dynamical Franz-Keldysh effect \cite{Lucchini916}, photocarrier
relaxation \cite{Zurch2017}, exciton dynamics \cite{Moulet1134}
as well as photocarrier injection \cite{Schlaepfer2018}.

Two-dimensional (2D) materials have been attracting great interest from
both fundamental and technological points of view because of their remarkable properties.
For example, graphene-like 2D materials such as hexagonal boron-nitride
(\textit{h}-BN) have been intensively studied for applications in optelectronic devices
and energy storage among others
\cite{RevModPhys.81.109,wang2017graphene,doi:10.1021/acs.chemrev.6b00558}.
Transition metal dichalcogenides (TMDs), another kind of 2D materials,
are also attracting attention because of their interesting properties 
deriving from strong spin-orbit coupling 
and a large direct-bandgap
\cite{Wang2012,Manzeli2017,doi:10.1021/acs.chemrev.6b00558}.
Thanks to recent development of laser technologies,
laser-induced ultrafast electron dynamics in such 2D materials has been
intensively studied both experimentally and theoretically 
\cite{Gierz2013,doi:10.1021/acs.jctc.6b00897,
PhysRevLett.117.277201,Shin2018,Sentef2015,Liu2017,PhysRevB.79.081406}.
For example, the high-order harmonic generation was recently studied
in 2D materials
\cite{Yoshikawa736,Liu2016,PhysRevB.95.085436,Tancogne-Dejeaneaao5207}.
Surprisingly, it was demonstrated that elliptically polarized light
can enhance the high-order harmonic generation in graphene \cite{Yoshikawa736}.

In order to fully understand such laser-induced ultrafast electron dynamics,
ATAS is one of the most desirable experimental techniques as 
it can resolve the dynamics with natural time-resolution.
While ATAS for solid-state materials provides a wealth of information on 
microscopic electron dynamics in materials,
the resulting experimental data are often very complicated due to
complex electronic structure and highly-nonlinear effects in the dynamics.
Therefore, it is hard to directly interpret the experimental results.
\textit{Ab-initio} simulation based on the time-dependent density functional
theory (TDDFT) \cite{PhysRevLett.52.997} is a powerful tool to describe 
such complex electron dynamics
and to provide microscopic insight into the phenomena.
Indeed, \textit{ab-initio} TDDFT simulations have been applied 
to ATAS experiments for solid-state systems
and played a significant role to construct the interpretation
\cite{Schultze1348,Lucchini916,Zurch2017,Schlaepfer2018}.

Likewise, experimental results of ATAS for 2D materials are
expected to be rather complicated.
Therefore, \textit{ab-initio} TDDFT simulations are a good candidate
to analyze ATAS for 2D materials.
To realize such simulation methods, 
we extend the TDDFT pump-probe simulation \cite{PhysRevB.89.064304} of
transient absorption spectroscopy to 2D materials.
To demonstrate the \textit{ab-initio} ATAS simulation, 
we employ monolayer \textit{h}-BN as an example of a 2D material
and investigate its transient optical property under intense infrared (IR)
laser pulses.
Furthermore, we analyze the obtained transient absorption spectra of \textit{h}-BN
with a 2D parabolic two-band model and clarify the microscopic mechanism
of the ultrafast modification of the optical property. 

This paper is organized as follows: In Sec.~\ref{sec:method}
we first describe the electron dynamics simulation based on TDDFT.
Then, we further describe the linear response calculation
and the pump-probe simulation to compute static and transient absorption properties,
respectively.
In Sec.~\ref{sec:atas} we apply the TDDFT pump-probe simulation to monolayer
\textit{h}-BN and investigate its transient optical properties.
Furthermore, we analyze the obtained transient absorption spectra with
a parabolic two-band model.
Finally, our findings are summarized in Sec.~\ref{sec:summary}.

%%%%%%%%%%%%%%%%%%%%%%%%%%%%%%%%%%%%%%%%%%
\section{Methods  \label{sec:method}}

In this section, we describe our theoretical and numerical methods 
to simulate ATAS for 2D materials.
As an example of 2D materials, we choose monolayer \textit{h}-BN in this work.

%====================================================================
\subsection{Electron dynamics simulation for periodic systems}

First, we briefly describe the first-principles
electron dynamics simulation based on 
TDDFT \cite{PhysRevLett.52.997}.
Details of the simulation are described elsewhere \cite{PhysRevB.62.7998}.
In the framework of TDDFT, electron dynamics in a periodic system is described by
the following time-dependent Kohn-Sham equation,
\be
i\hbar\frac{\partial}{\partial t} u_{b\vec k}(\vec r,t)
&=&\left [ 
\frac{1}{2m_e} \left \{ \vec p + \hbar \vec k + \frac{e}{c}\vec A(t)
\right \}^2 + \hat v_{ion} + v_H(\vec r,t) + v_{XC}(\vec r,t)
\right ]u_{b\vec k}(\vec r,t) \nonumber \\
&=& H_{KS,\vec k}(t) u_{b\vec k}(\vec r,t), 
\label{eq:tdks}
\ee
where $u_{b\vec k}(\vec r,t)$ is the Kohn-Sham Bloch wavefunction, 
$b$ is a band index, and $\vec k$ is the Bloch wave vector. The electron-ion interaction
is described by the ionic potential $\hat v_{ion}$, while the electron-electron interaction
is described by the combination of the Hartree potential $v_H(\vec r,t)$ and 
the exchange-correlation potential $v_{XC}(\vec r,t)$.
The Hartree potential satisfies the Poisson equation, 
$\nabla^2 v_H(\vec r,t) = 4e^2\pi \rho(\vec r,t)$, where the electron density
$\rho(\vec r,t)$ is evaluated as
\be
\rho(\vec r, t) = \sum_{b \vec k} n_{b \vec k} \left | 
u_{b \vec k}(\vec r,t)
\right |^2,
\ee
with occupation factors, $n_{b\vec k}$.

The exchange-correlation potential $v_{XC}(\vec r,t)$ 
is a functional of the electron density.
Because its exact form is unknown, one has to approximate it.
In this work, we employ the adiabatic
local-density approximation \cite{PhysRevB.45.13244} (ALDA), where the exchange-correlation
potential depends on the electron density locally in space and time.
External laser fields are described by spatially-uniform 
time-varying vector potentials $\vec A(t)$ in the dipole approximation.

To describe monolayer \textit{h}-BN, we employ three-dimensional
periodic boundary conditions with a hexagonal lattice.
The real-space atomic configuration in the plane is shown in
Fig.~\ref{fig:current_lin}~(a). In this work, $y$-axis is taken to be parallel
to a B-N bond, while $x$-axis is taken to be perpendicular to it.
The lattice constant of the hexagonal lattice is set to $a=4.76$~a.u.
The out-of-plane direction is assigned to the $z$-axis, and 
the period of that direction is set to $L=50$~a.u., which is large enough to avoid
spurious inter-layer interactions.

We discretize the hexagonal lattice into three-dimensional real-space 
grid points with a grid spacing about 0.36~a.u.
We further discretize the two-dimensional first Brillouin zone 
of \textit{h}-BN into $32\times32$ \textit{k}-points.

In this work, dynamics of core electrons in the $1s$ shell of Boron and Nitrogen atoms 
are not expected to be significant for transient absorption. 
Therefore, we treat explicitly only valence electrons
while the core electrons are frozen. For this purpose,
we employ the Hartwigsen-Goedecker-Hutter pseudopotentials \cite{PhysRevB.58.3641}
to describe interactions between valence electrons and ionic cores.

All the density functional theory (DFT) and TDDFT calculations in this work are carried out
with the \textit{Octopus} code \cite{C5CP00351B}.

%\begin{figure}[H]
%\centering
%\includegraphics[width=0.7\columnwidth]{hbn_structure.pdf}
%\caption{\label{fig:hbn_structure} Atomic configuration of the monolayer \textit{h}-BN.}
%\end{figure}  

%====================================================================
\subsection{Optical property from linear response calculation}

\begin{figure}[H]
\centering
\includegraphics[width=0.34\columnwidth]{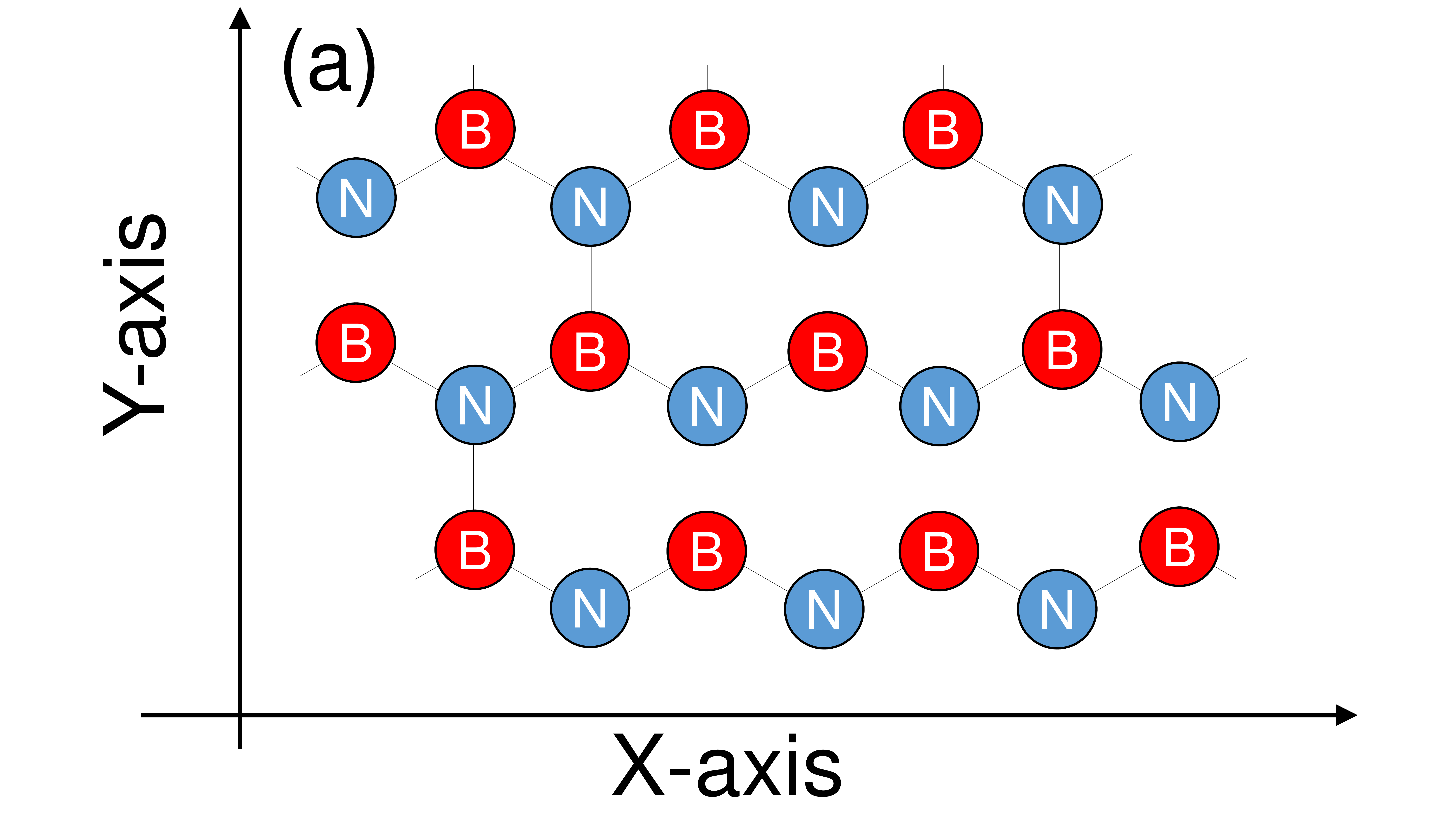}
\includegraphics[width=0.65\columnwidth]{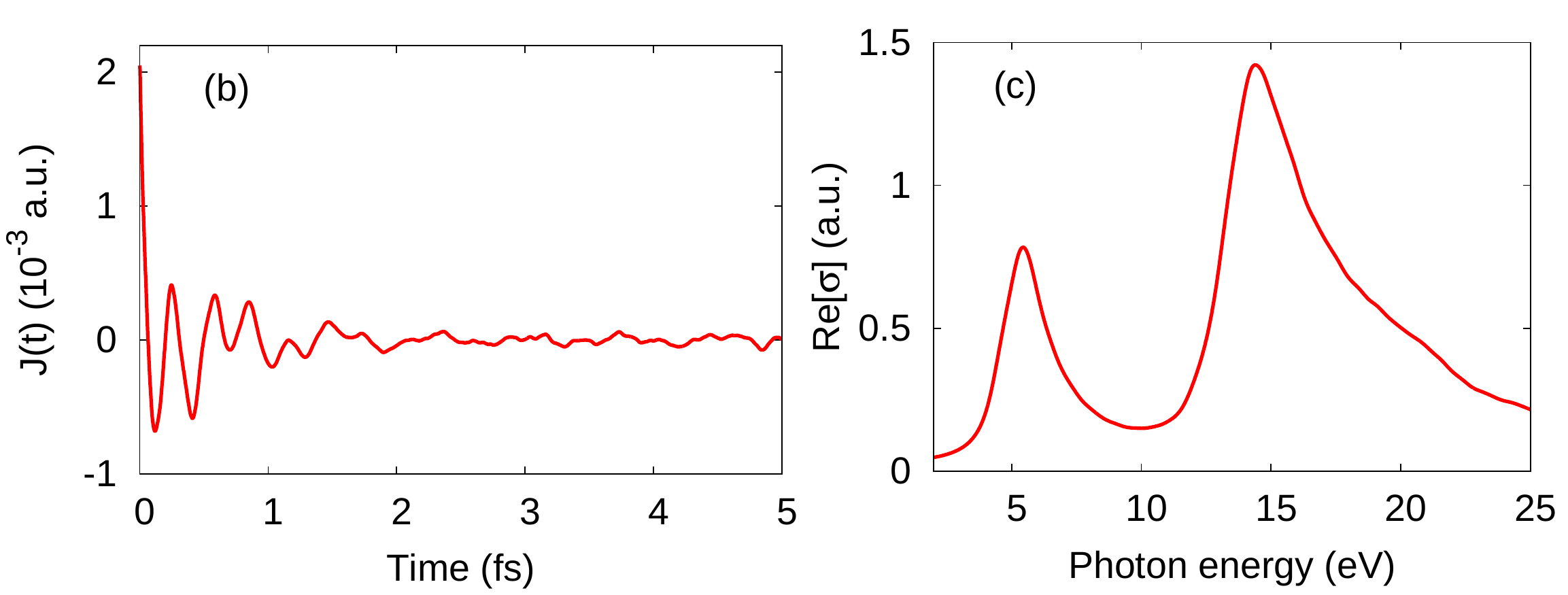}
\caption{\label{fig:current_lin} 
(a) Atomic configuration of monolayer \textit{h}-BN.
(b) Current density in time domain after the impulsive distortion at $t=0$.
(c) Real-part of surface conductivity, $\Re \left[\sigma(\omega) \right]$, 
of \textit{h}-BN, calculated by TDDFT with ALDA.}
\end{figure}

Here, we revisit the linear response calculation in time domain 
to obtain optical properties of 2D materials.
For this purpose, we consider electron dynamics in monolayer \textit{h}-BN
under an impulsive perturbation, $\vec E(t)=\vec e_x k_0 \delta(t)$.
The polarization of the perturbation is set along the $x$-direction in 
Fig.~\ref{fig:current_lin}~(a), which is an in-plane direction.
The strength of the perturbation $k_0$ is set to 0.005~a.u.,
which is weak enough to only create excitations on the linear response regime.

In this context, one of the most useful results of TDDFT calculations 
for 2D materials is the surface density of the electric current,
\be
\vec J(t) = -\frac{e}{S}\sum_{b\vec k} n_{b\vec k}
\int_{cell} d\vec r u^*_{b\vec k}(\vec r,t) \vec v_{\vec k}(t)
u_{b\vec k}(\vec r,t),
\label{eq:current}
\ee
where $S$ is the area of the \textit{h}-BN sheet in the simulation cell, and 
$\vec v_{\vec k}$ is the velocity operator:
\be
\vec v_{\vec k}(t)= \frac{\left [ \vec r, H_{KS,\vec k}(t) \right ]}{i\hbar} 
= \frac{1}{m_e} \left [\vec p + \hbar \vec k + \frac{e}{c}\vec A(t)  \right ]
+ \frac{\left[ \vec r,\hat v_{ion} \right]}{i\hbar}.
\ee
Note that the velocity operator contains a contribution from 
the ionic potential $\hat v_{ion}$
because of the nonlocality of the pseudopotential,
which is not commutative with the position operator $\vec r$ \cite{PhysRevB.62.7998}.

Figure~\ref{fig:current_lin}~(b) shows the surface current density, $J(t)$,
for the $x$-direction as a function of time.
Since the impulsive distortion is applied at $t=0$, the current is suddenly
induced at that time. Then, the induced current shows oscillatory behavior with damping.

One can extract the optical conductivity from the computed current in time domain.
The diagonal element of the conductivity can be evaluated as
\be
\sigma(\omega) = \frac{\int dt J(t) e^{i\omega t- \gamma t}}
{\int dt E(t) e^{i\omega t- \gamma t}},
\label{eq:sigma_lin}
\ee
where $J(t)$ and $E(t)$ are the electric current and the electric field
for the polarization direction of the impulsive distortion, respectively.
To reduce numerical error due to the finite time propagation in the Fourier transform
of Eq.~(\ref{eq:sigma_lin}),
we employed a damping factor $\gamma$ of 0.5~eV$/\hbar$.

Figure~\ref{fig:current_lin}~(c) shows the real-part of the optical conductivity
of \textit{h}-BN for the in-plane direction, which is computed from the current
in Fig.~\ref{fig:current_lin}~(b). The computed conductivity shows two peaks at around
5~eV and 15~eV. This double-peak structure is consistent with the previous theoretical work
based on the first-principles simulation \cite{BEIRANVAND2015190}.
Because the real-part of the conductivity is closely related to
the optical absorption, we treat it as a direct measure of the photo-absorption
in this work.
We note that the present calculation with ALDA does not properly describe
the excitonic effect which is significant in the optical absorption of 
\textit{h}-BN around the optical gap \cite{PhysRevLett.96.126104,PhysRevLett.116.066803}.
However, as will be shown later, we focus on transitions to conduction states with
relatively high energy, where the excitonic effect is expected to be less important.
Therefore, we simply ignore the excitonic contribution in this work.

%==============================================================================
\subsection{Transient optical properties with pump-probe simulations}

Now we describe the pump-probe simulation to investigate
transient optical properties. 
Details of the numerical pump-probe simulation are described elsewhere
\cite{PhysRevB.89.064304}.

To compute transient optical properties of materials,
we consider electron dynamics under pump $\vec E_{pump}(t)$ and 
probe $\vec E_{probe}(t)$ electric fields.
Solving the time-dependent Kohn-Sham equation~(\ref{eq:tdks})
under the presence of both the pump and probe fields,
one can simulate the electron dynamics induced by the fields.
Furthermore, one can calculate the induced electric current by using Eq.~(\ref{eq:current});
we shall call the current under both the pump and probe pulses
\textit{pump-probe current}, $\vec J_{pump-probe}(t)$.
Additionally, one can simulate electron dynamics under only the pump pulse,
and compute the current: we shall call the current induced solely by the pump pulse
\textit{pump current}, $\vec J_{pump}(t)$.
To extract the current induced by the probe pulse under the presence of the pump, 
we define the
\textit{probe current}, $\vec J_{probe}(t)$, as
\be
\vec J_{probe}(t) \equiv
\vec J_{pump-probe}(t) - \vec J_{pump}(t).
\ee

Finally, referring to the above linear response equation,
the transient conductivity can be evaluated as
\be
\sigma^T(\omega,T_{probe}) = 
\frac{\int dt J_{probe}(t) e^{i\omega t- \gamma t}}
{\int dt E_{probe}(t) e^{i\omega t- \gamma t}},
\label{eq:sigma_t}
\ee
where $T_{probe}$ is the central time of the probe pulse.
In contrast to the linear response calculation in equilibrium,
the transient conductivity under the pump pulse
depends also on the central time of the probe pulse since
the time translation symmetry is broken by the pump.
Thus, one may investigate dynamics of transient optical properties
in time domain by changing the time delay between pump and probe pulses.

To practically carry out the pump-probe simulations,
we employ a pump vector potential of the following form;
\be
\vec A_{pump}(t) &=& -\frac{c\vec E_{pump}}{\omega_{pump}}
\cos^2\left (\pi \frac{t}{T_{pump}} \right ) \sin \left (\omega_{pump} t\right)
\ee
in the domain $(-T_{pump}/2 <t<T_{pump}/2)$ and zero outside of it.
Here, $\vec E_{pump}$ is the peak electric field, $\omega_{pump}$ is 
the mean frequency, and $T_{pump}$ is the full duration of the pump pulse.

To perform the pump-probe simulation, we also employ a similar form for the probe
vector potential;
\be
\vec A_{probe}(t) &=& -\frac{c\vec E_{probe}}{\omega_{probe}}
\cos^4\left (\pi \frac{t-T_{delay}}{T_{probe}} \right ) 
\sin \left [\omega_{pump} \left (t-T_{delay} \right )\right]
\ee
in the domain $(-T_{probe}/2 <t-T_{delay}<T_{probe}/2)$ and zero outside of it.
Here, $\vec E_{probe}$ is the peak electric field, $\omega_{probe}$ is 
the mean frequency, $T_{probe}$ is the full duration of the probe pulse.
The time delay between pump and probe pulses is expressed by $T_{delay}$.

%================================================================================
\section{Attosecond transient absorption of monolayer \textit{h}-BN 
\label{sec:atas}}

In this section, we investigate transient optical properties of monolayer
\textit{h}-BN under intense IR pulses.
For this purpose, we employ the TDDFT pump-probe approach explained in
Sec.~\ref{sec:method}.

For the pump-probe simulation, we first set the polarization direction 
of both the pump and probe pulses to the $x$-direction in
Fig.~\ref{fig:current_lin}~(a).
Thus, the polarization direction of the pump and probe pulses is
perpendicular to a B-N bond. Note that \textit{h}-BN has 
inversion symmetry along this direction.
For the pump pulse, the peak field strength $|\vec E_{pump}|$ is set to 
$8.7\times 10^8$~V/m,
the mean frequency $\omega_{pump}$ is set to 1.55~eV$/\hbar$, 
and the full pulse duration $T_{pump}$ is set to 20~fs.
The corresponding full width at half maximum (FWHM) of the pump pulse is
about 7.3~fs.
For the probe pulse, $|\vec E_{probe}|$ is set to $8.7\times 10^7$~V/m,
$\omega_{probe}$ is set to 15~eV$/\hbar$, and $T_{probe}$ is set to 1~fs.
The corresponding FWHM of the probe pulse is about 260~as.
Under these conditions, we repeat the TDDFT pump-probe simulations
by changing the pump-probe time-delay, $T_{delay}$.

\begin{figure}[H]
\centering
\includegraphics[width=0.8\columnwidth]{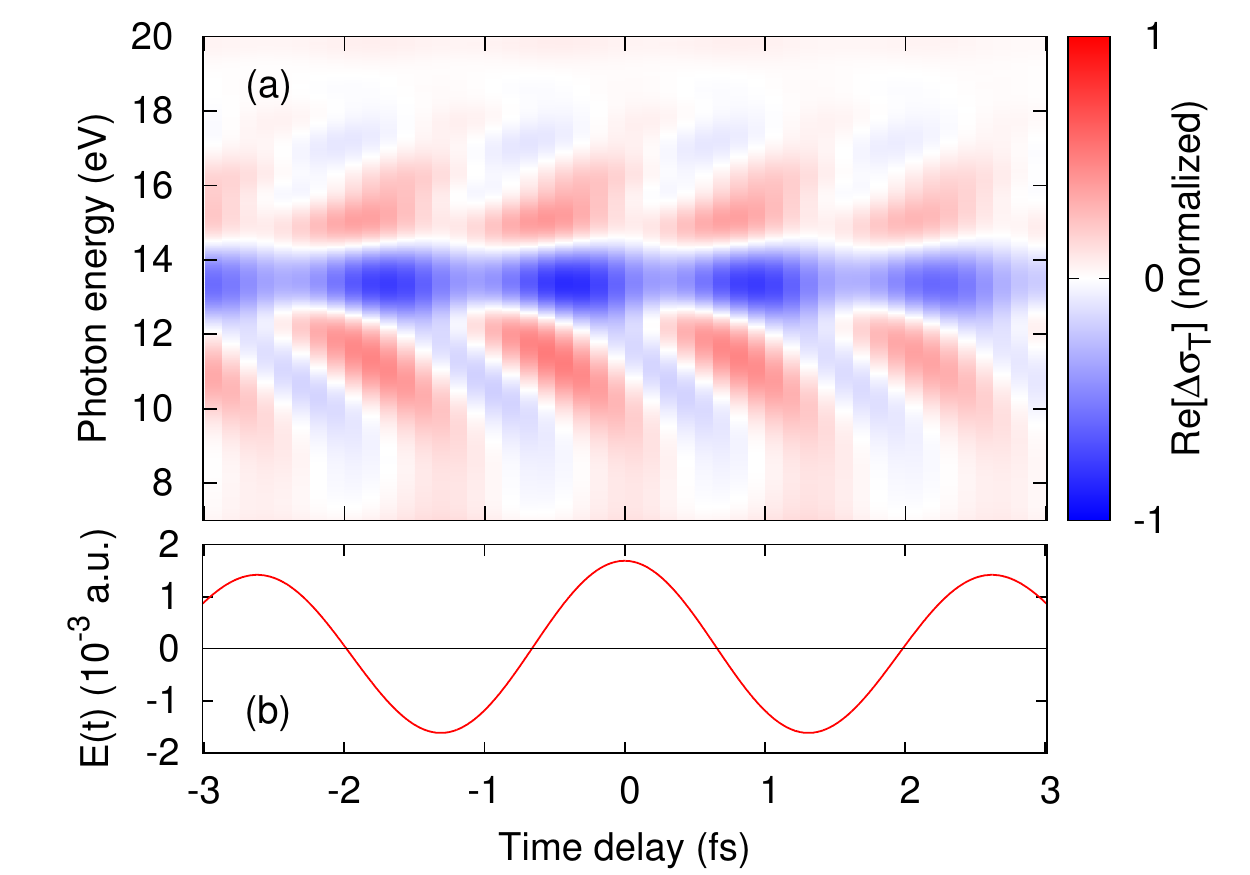}
\caption{\label{fig:tr_sigma} (a) Real-part of the transient conductivity 
of monolayer \textit{h}-BN under the IR pump pulse for the $x$-direction
in Fig.~\ref{fig:current_lin}~(a).
(b) The time-profile of the applied pump electric field.
The result is computed by TDDFT with ALDA.
}
\end{figure}

Figure~\ref{fig:tr_sigma}~(a) shows the real-part of the transient conductivity
$\sigma_T(\omega,T_{delay})$ computed by the TDDFT pump-probe simulations.
The time-profile of the applied pump electric field 
is also shown in Fig.~\ref{fig:tr_sigma}~(b).
In Fig.~\ref{fig:tr_sigma}~(a), one sees an oscillatory feature with a V-shaped
energy dispersion around 15~eV, which we shall call \textit{fishbone} structure.
As seen from the figure, the frequency of the oscillation in time domain is twice of 
the pump frequency $\omega_{pump}$.
The $2\omega_{pump}$-oscillation is a direct consequence of the inversion symmetry
of the material in the pump-probe direction, which forbids even-order nonlinear responses,
because materials with inversion symmetry must show the same response
regardless of the sign of the electric field along a symmetry direction.
Similar features have been observed in attosecond transient absorption
spectroscopy for semiconductors \cite{Lucchini916,Schlaepfer2018}, and
transient terahertz absorption spectroscopy for GaAs quantum wells
\cite{PhysRevLett.117.277402}.
These features of bulk materials have been discussed on the basis of
the pump-induced intraband transitions and the dynamical Franz-Keldysh effect 
\cite{PhysRevLett.76.4576,PhysRevB.93.045124}.
The dynamical Franz-Keldysh effect is a modification of optical properties of
solids under the influence of oscillatory electric fields.
It originates from the field-induced intraband transitions via the modification
of the dynamical phase factor.

%In order to understand the microscopic origin of the fishbone structure
%in Fig.~\ref{fig:tr_sigma},
%we investigate the electronic structure of the monolayer \textit{h}-BN.
%Figure~\ref{fig:band_map_hbn} shows the band structure of
%the monolayer \textit{h}-BN, computed by the DFT calculation. 
%The valence bands are described by the red lines, while
%the conduction bands are described by the blue lines.
%One can see that vertical transitions from the highest valence band to higher energy
%conduction bands occur with the excitation energy of 15~eV around
%the $M$ point (see the black arrow in Fig.~\ref{fig:band_map_hbn}).
%Since these transition energies are close to the energy range of
%the fishbone structure in Fig.~\ref{fig:tr_sigma},
%vertical transitions around the $M$ point are expected 
%to play a significant role.
%
%\begin{figure}[H]
%\centering
%\includegraphics[width=0.60\columnwidth]{band_map_hbn}
%\caption{\label{fig:band_map_hbn} Electronic structure of the monolayer \textit{h}-BN
%computed by DFT calculation with the LDA \cite{PhysRevB.45.13244}.
%The energy of the valence top is set to zero.
%}
%\end{figure} 

In order to understand the microscopic origin of the fishbone structure
in Fig.~\ref{fig:tr_sigma},
we investigate energies of photo-excited electron-hole pairs
based on the independent particle approximation.
For this purpose, we compute the vertical excitation energies
from the highest valence band to the conduction bands.
Note that in the independent particle approximation 
the transition energy from one state to another is identical to
the difference of the single-particle energies of those states, 
$\epsilon_{c\vec k}-\epsilon_{v\vec k}$.

\begin{figure}[H]
\centering
\includegraphics[width=0.80\columnwidth]{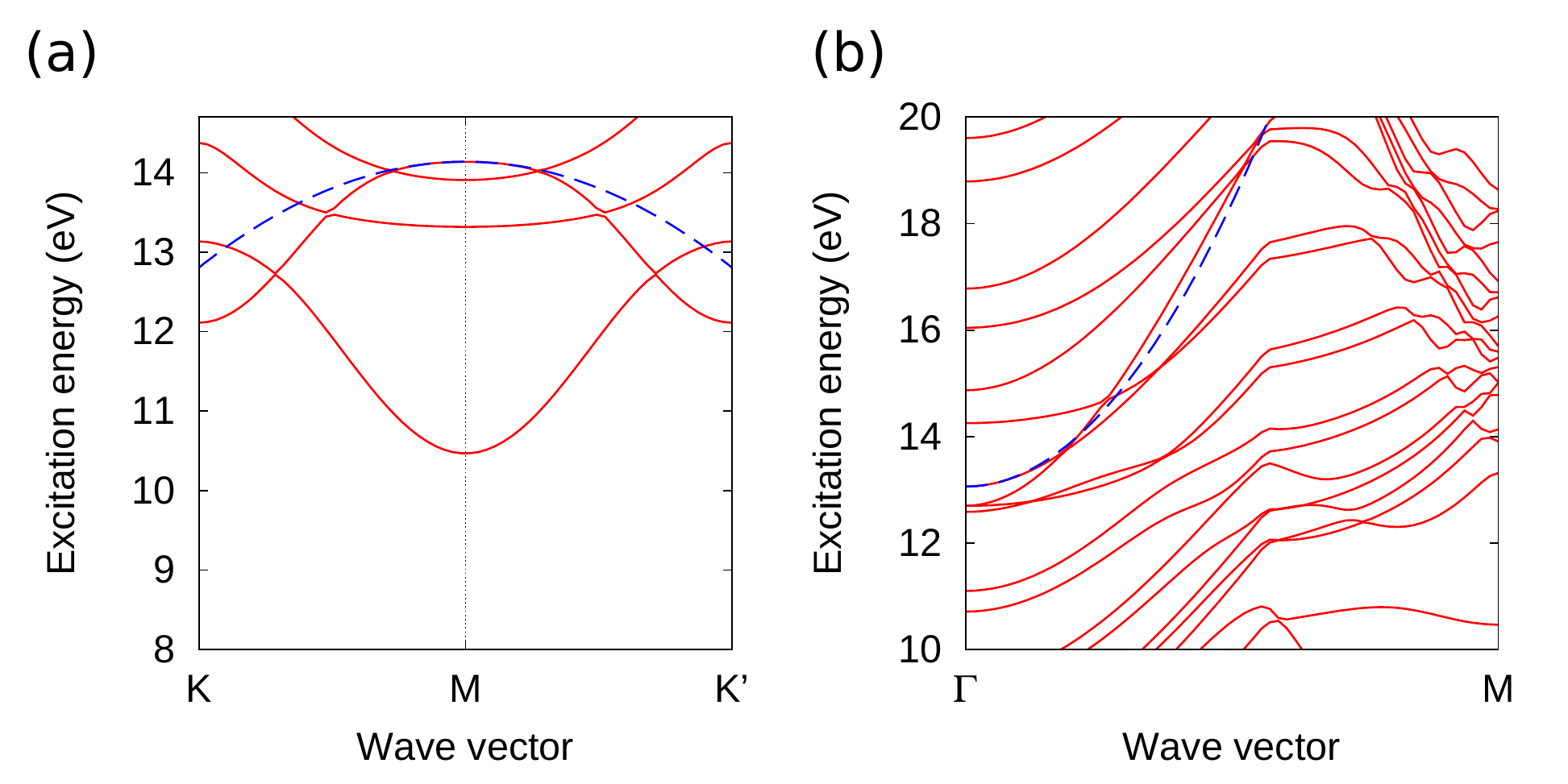}
\caption{\label{fig:ph_map_hbn} Vertical transition energies from
the highest valence band to conduction bands in monolayer \textit{h}-BN.
The transition energies are computed by
the difference of the Kohn-Sham eigenenergies of the highest valence band
and conduction bands, $\epsilon_{c\vec k}-\epsilon_{v\vec k}$,
corresponding to excitations in the independent particle approximation.
}
\end{figure}

Figure~\ref{fig:ph_map_hbn} shows the computed vertical transition energies from the highest
valence band to the conduction bands at each $k$ point 
along high symmetry paths through Brillouin zone. 
Transitions around the $M$ point are described in Fig.~\ref{fig:ph_map_hbn}~(a),
while those around the $\Gamma$ point are described in 
Fig.~\ref{fig:ph_map_hbn}~(b).
In Fig.~\ref{fig:ph_map_hbn}~(a), one can see a negative electron-hole mass band 
around 14~eV at the $M$ point.
The negative-mass band is fitted by
a parabola (blue-dashed line) to extract band parameters. 
The extracted reduced electron-hole mass 
at the $M$-point is $-3.3m_e$.

Furthermore, we numerically confirmed that all the other vertical transitions 
in Fig.~\ref{fig:ph_map_hbn}~(a) at the $M$ point are dipole forbidden.
Therefore, the negative-mass band is expected to have a dominant contribution
to the transient absorption spectrum in Fig.~\ref{fig:tr_sigma}.

To elucidate the contribution from the negative-mass band at the $M$ point, 
we compute the transient absorption spectrum with a parabolic two-band model.
The parabolic two-band model is described in details elsewhere \cite{PhysRevB.98.035202}.

In the parabolic two-band model, the electron dynamics at each $k$ point in 
the Brillouin zone is described by the following Schr\"odinger equation
for two-dimensional state vectors,
\be
i\hbar
\frac{d}{dt}
\left(
    \begin{array}{c}
      c_{v\vec k}(t)   \\
      c_{c\vec k}(t)
    \end{array}
  \right)
=
\left(
    \begin{array}{cc}
      0 & h_{vc,\vec k}(t)  \\
      h^*_{vc, \vec k}(t) & 0
    \end{array}
  \right)
\left(
    \begin{array}{c}
      c_{v\vec k}(t)   \\
      c_{c\vec k}(t)
    \end{array}
  \right),
\nonumber \\
\label{eq:schrodinger-2band}
\ee
where the off-diagonal matrix element of the Hamiltonian is given by
\be
h_{vc,\vec k}(t) = \frac{i \vec p_{vc,\vec K(t)} \cdot \vec E_{probe}(t) }
{\Delta \epsilon_{cv, \vec K(t)} }\frac{e \hbar}{m}
e^{\frac{1}{i\hbar}\int^t dt' 
\Delta \epsilon_{cv, \vec K(t')}},
\ee
where $\Delta \epsilon_{cv, \vec k}$ is the transition energy between
the valence and the conduction states at $\vec k$,
and  $\vec p_{vc,\vec k}$ is the transition momentum.
In the Hamiltonian matrix, the Bloch wavevector is shifted 
based on the acceleration theorem, $\vec K(t)=\vec k + e\vec A_{pump}(t)/\hbar c$,
by the pump pulse \cite{PhysRev.57.184}.

Here, in order to elucidate the role of the pump-induced intraband transitions,
only the probe electric field $\vec E_{probe}(t)$ is coupled to the transition
momentum $\vec p_{vc}$, while only the pump vector potential $\vec A_{pump}(t)$
is taken into account in the acceleration theorem. Thus, the probe pulse
induces only the interband transitions, while the pump pulse is responsible for
intraband transitions.

For this model, we approximate the transition energy by the parabolic band as
\be
\Delta \epsilon_{cv,\vec k} = \epsilon_g + \frac{\hbar ^2}{2\mu_{eh}}k^2,
\ee
where $\epsilon_g$ is the band gap, and $\mu_{eh}$ is the reduced mass
of an electron-hole pair.
To describe the negative-mass band at the $M$ point,
the reduced mass $\mu_{eh}$ is set to $\mu_M = -3.3m_e$, 
and the band gap $\epsilon_g$ is set to $\epsilon_{g,M}=14.1$~eV.
These parameters are extracted from the parabolic fit shown as the blue-dashed line in
in Fig.~\ref{fig:ph_map_hbn}.

The negative-mass band causes an artificial crossing of the conduction and
the valence bands. To eliminate the effect from the artificial band crossing of the model,
we truncate the transition momentum matrix as follows
\be
p_{vc,\vec k} = 
\left\{
  \begin{array}{@{}ll@{}}
    p_{vc}, & (\epsilon_{g,M} \leq \Delta \epsilon_{cv,\vec k} ) \\
    p_{vc} \sin^2\left ( \frac{\pi \Delta \epsilon_{cv,\vec k}}{2\epsilon_{g} }\right )
, & (0<\Delta \epsilon_{cv,\vec k} <\epsilon_{g}) \\
    0, & \text{otherwise}
  \end{array}\right.
\ee
Here, the transition momentum at the gap $p_{vc}$ is set to 0.1~a.u.
Note that the value of the transition momentum does not affect the structure of 
absorption spectra as long as the field strength of the probe pulse is weak enough
to satisfy the linear response condition.

We apply a pump-probe simulation to the parabolic two-band model and compute
the transient absorption spectrum of the negative-mass band.
Since the parabolic-band approximation is an appropriate approximation of 
the electronic structure only around the band edge, 
we truncate the signal to reduce the contribution from inappropriate regions
by applying the following attenuation function to the transient conductivity
\be
w_M(\omega) = 
\left\{
  \begin{array}{@{}ll@{}}
    1, & (\omega  > \epsilon_{g,M}/\hbar) \\
    \exp \left[ -\frac{1}{2\sigma^2} (\hbar \omega -\epsilon_{g,M} )^2\right ], & \text{otherwise}
  \end{array}\right.
\ee
Here, $\epsilon_{g,M}$ is the corresponding band gap at the $M$-point, 
which is 14.1~eV. 
We set the width of the attenuation $\sigma$ to 1~eV.

\begin{figure}[H]
\centering
\includegraphics[width=0.8\columnwidth]{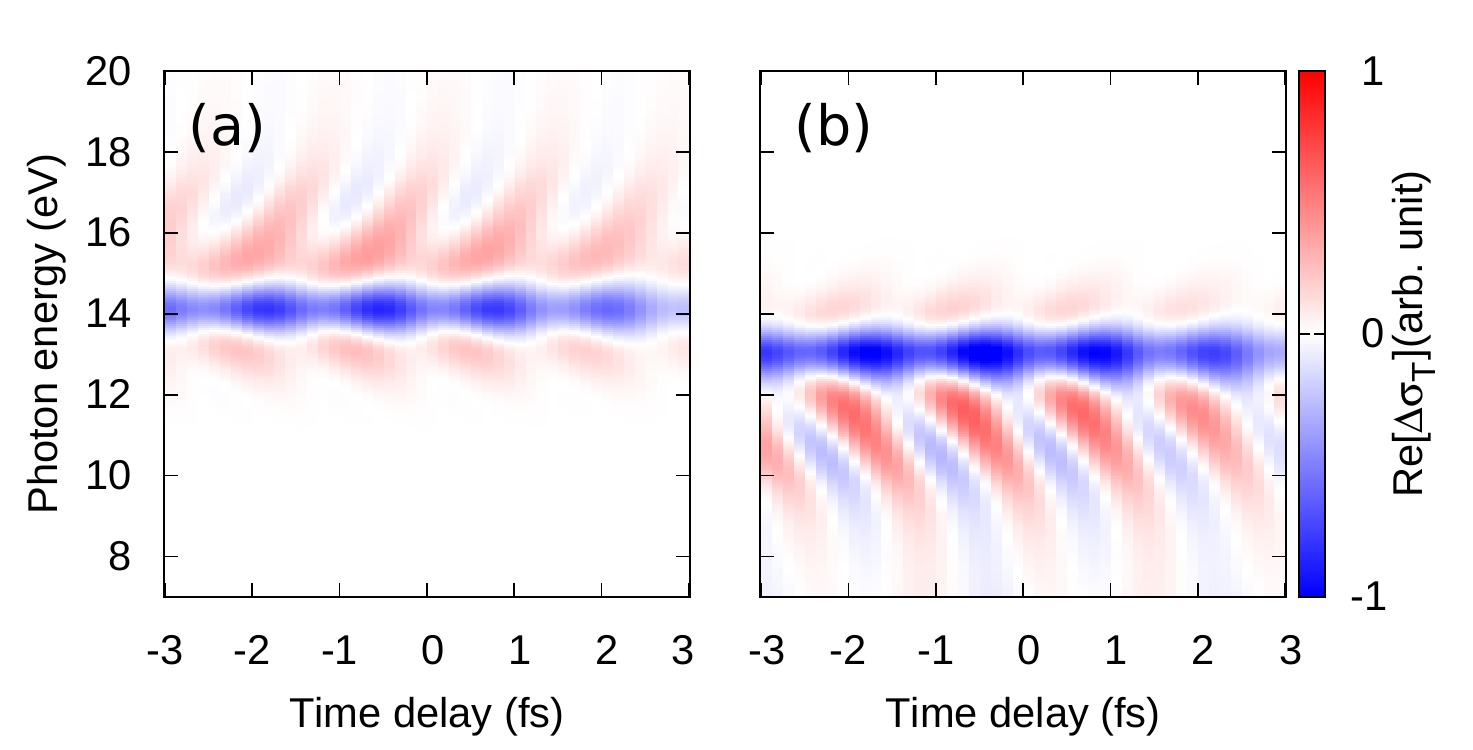}
\caption{\label{fig:tr_sigma_2band_mg_decomp} 
Transient surface conductivities of parabolic two-band models under the femtosecond 
IR pulse. The panels~(a) and (b) show the result of 
the parabolic band shown as the blue-dashed line 
in Fig.~\ref{fig:ph_map_hbn}(a) and (b), respectively.}
\end{figure}

Figure~\ref{fig:tr_sigma_2band_mg_decomp}~(a) shows the truncated transient conductivity,
$\sigma^T(\omega,T_{probe})w_M(\omega)$, of the negative-mass band at the $M$ point.
One may see that the transient absorption spectrum of the negative-mass parabolic band model
in Fig.~\ref{fig:tr_sigma_2band_mg_decomp}~(a) nicely reproduces the upper-half part of 
the transient absorption spectrum of the full \textit{ab-initio} simulation 
in Fig.~\ref{fig:tr_sigma}.
This fact indicates that the upper-half part of the fishbone structure
originates from this specific band around the $M$ point, which 
has the anomalous band dispersion with the negative reduced electron-hole mass.

To reveal the other half contribution in the transient absorption spectrum of \textit{h}-BN,
we investigate vertical transitions around the $\Gamma$ point.
Figure~\ref{fig:ph_map_hbn}~(b) shows the vertical transition energies from
the highest valence band to conduction bands around the $\Gamma$ point.
Among those of transitions around the energy of 13~eV at the $\Gamma$-point,
we pick up a single band, which has the largest transition momentum for the polarization
direction of the pump and probe pulses. 
The chosen band is fitted by a parabola that is shown as the blue dashed
line in Fig.~\ref{fig:ph_map_hbn}~(b).
The extracted band parameters are as follows: the band gap $\epsilon_g$ is 13.1~eV, and
the reduced electron-hole mass $\mu_{eh}$ is $1.1m_e$.
Employing the parabolic two-band model with these extracted parameters,
we computed the transient optical conductivity based on the pump-probe simulation.
As explained above, the parabolic fit is only accurate around the band edge.
Therefore, we apply the following attenuation function to the transient optical
conductivity to reduce the influence from irrelevant energy regions;
\be
w_{\Gamma}(\omega) = 
\left\{
  \begin{array}{@{}ll@{}}
    1, & (\omega  < \epsilon_{g,\Gamma}/\hbar) \\
    \exp \left[ -\frac{1}{2\sigma^2} (\hbar \omega -\epsilon_{g,\Gamma} )^2\right ], & \text{otherwise}
  \end{array}\right.
\ee
Here, $\epsilon_{g,\Gamma}$ is the corresponding band gap, which is 13.1~eV.

Figure~\ref{fig:tr_sigma_2band_mg_decomp}~(b) shows the truncated transient conductivity,
$\sigma^T(\omega,T_{probe})w_{\Gamma}(\omega)$, with the parabolic two-band model
of the $\Gamma$ point. One can clearly see that the result of the parabolic
two-band model in Fig.~\ref{fig:tr_sigma_2band_mg_decomp}~(b) 
nicely reproduces the half bottom part of 
the full \textit{ab-initio} simulation result in Fig.~\ref{fig:tr_sigma}.

\begin{figure}[H]
\centering
\includegraphics[width=0.8\columnwidth]{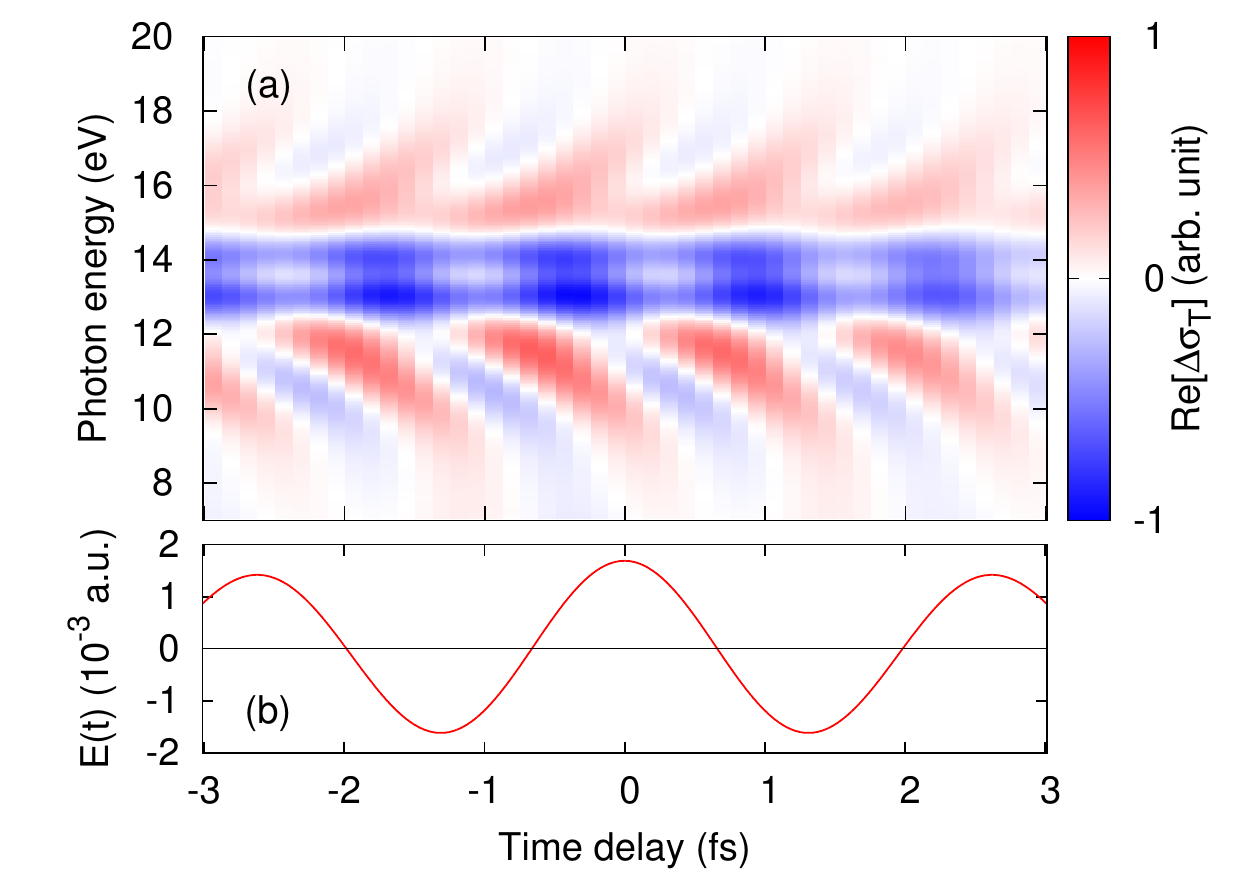}
\caption{\label{fig:tr_sigma_2band} 
(a)Transient surface conductivities of the combined two-band models at
$M$ and $\Gamma$ points. This result is nothing but the summation
of Figs.~\ref{fig:tr_sigma_2band_mg_decomp}~(a) and (b).
(b) The time-profile of the applied pump electric field.
}
\end{figure}

To provide the complete description, we combine 
the results of the two-band models for the $M$ and $\Gamma$ points
by simply summing the transient conductivities 
in Figs.~\ref{fig:tr_sigma_2band_mg_decomp}~(a) and (b).
The combined transient conductivity is shown in Fig.~\ref{fig:tr_sigma_2band}.
One sees that the combined result nicely reproduces the full \textit{ab-initio}
result in the whole investigated photon-energy range.
Therefore, we can conclude that the fishbone structure
in the transient absorption spectrum of monolayer \textit{h}-BN
consists of two major contributions:
One is the contribution from the negative-mass electron-hole band at the $M$ point, 
accounting for the upper half of the fishbone structure. The other is the contribution
from the positive-mass band at the $\Gamma$ point, inducing
the lower half of the fishbone structure.

Since the pump-induced interband transitions are completely omitted
in the model by construction, the transient absorption of the two-band model
in Fig.~\ref{fig:tr_sigma_2band} results only from the IR-induced intraband transitions.
Therefore, we clearly demonstrated that the IR-induced intraband transitions
play crucial roles and the transient absorption is dominated
by the dynamical Franz-Keldysh effect even in this 2D material.

Furthermore, based on the decomposition of the transient absorption spectrum
in Fig.~\ref{fig:tr_sigma_2band} into the contributions
from each band in Figs.~\ref{fig:tr_sigma_2band_mg_decomp}~(a) and (b),
we found that the upper-half part of the transient absorption spectrum
of monolayer \textit{h}-BN originates from the dynamical Franz-Keldysh effect 
of a specific band around the $M$ point
with the anomalous band dispersion, while the lower-half part
originates from the conventional dynamical Franz-Keldysh effect with the positive-mass
electron-hole band.
As intraband motion is often regarded as semi-classical motion of a free particle,
the predominant role of intraband motion in the anomalous dispersion
has the potential to induce further interesting dynamics in 
the 2D material.

Finally, we report the transient optical property of monolayer \textit{h}-BN
for a different polarization direction. 
Here, we set the polarization direction of the pump and probe pulses to 
the $y$-axis in Fig.~\ref{eq:sigma_t}. 
Thus, the polarization direction is parallel to a B-N bond.
For this direction, the system does not have the spatial inversion symmetry and 
it may have even order nonlinear responses.
To investigate transient optical properties, we perform
pump-probe simulation with the following parameters.
For the pump pulse, the peak field strength $|\vec E_{pump}|$ is set to
$8.7\times10^{7}$~V/m,
the mean frequency $\omega_{pump}$ is set to 1.55~eV$/\hbar$, 
and the full pulse-duration $T_{pump}$ is set to 20~fs.
For the probe pulse, $|\vec E_{probe}|$ is set to $8.7\times10^{7}$~V/m,
$\omega_{probe}$ is set to 15~eV$/\hbar$, and $T_{probe}$ is set to 1~fs.
Under these conditions, we repeat the TDDFT pump-probe simulations
by changing the pump-probe time-delay, $T_{delay}$.

\begin{figure}[H]
\centering
\includegraphics[width=0.8\columnwidth]{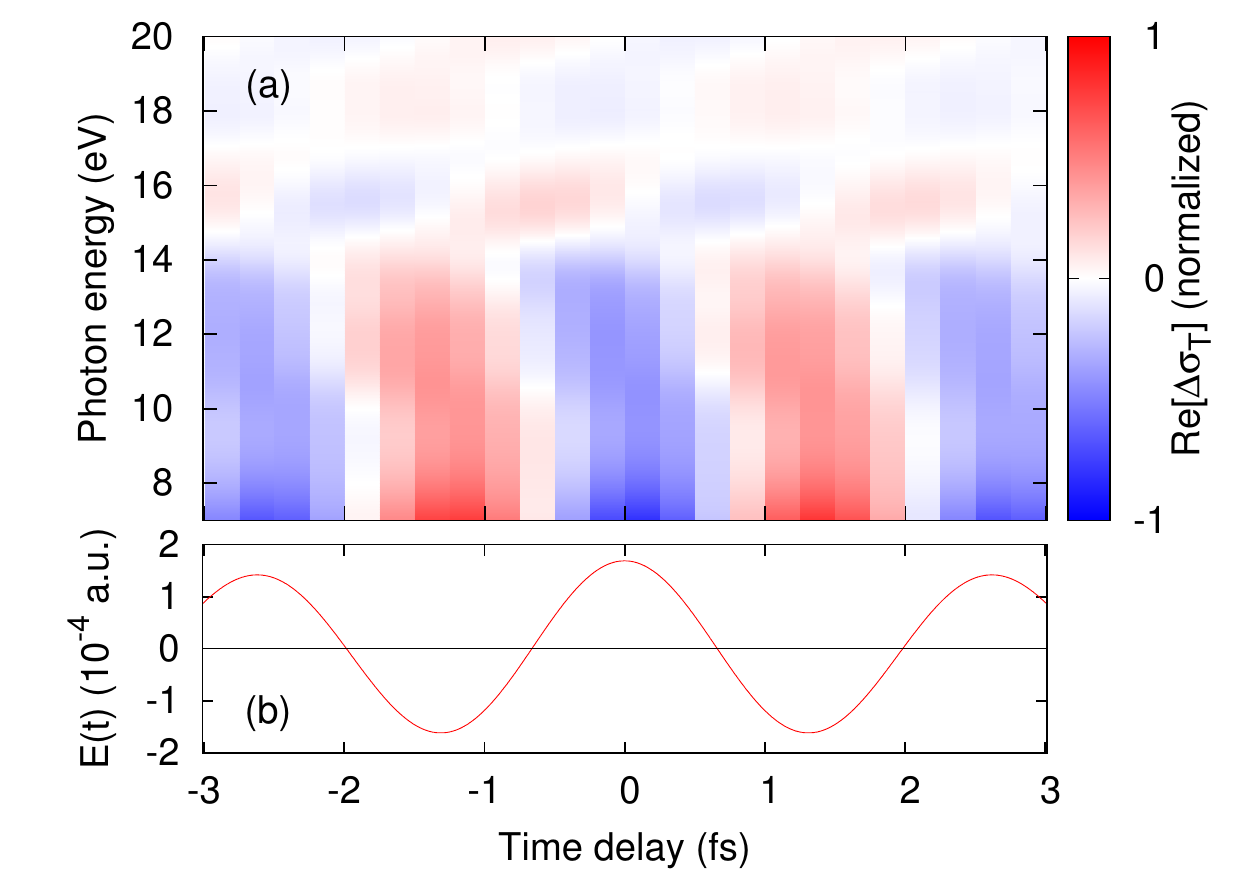}
\caption{\label{fig:tr_sigma_y} (a) Real-part of the transient conductivity 
of monolayer \textit{h}-BN under the IR pump pulse for the $y$-direction
in Fig.~\ref{fig:current_lin}~(a).
(b) The time-profile of the applied pump electric field.
The result is computed by TDDFT with ALDA.
}
\end{figure}

Figure~\ref{fig:tr_sigma_y}~(a) shows the computed transient conductivity
with the TDDFT pump-probe simulation,
and Fig.~\ref{fig:tr_sigma_y}~(b) shows the time profile of the applied pump
electric field. 
In contrast to the $2\omega_{pump}$-oscillation
in Fig.~\ref{fig:tr_sigma}, which was the consequence of the inversion symmetry,
the transient conductivity 
shows an oscillatory feature with the frequency of $\omega_{pump}$ in time domain.
The emergence of the $\omega_{pump}$-oscillation instead of 
the $2\omega_{pump}$-oscillation 
is due to the even-order nonlinear responses, which are forbidden
for the $x$-direction due to inversion symmetry.

Despite of the qualitatively different time-domain structures between
Fig.~\ref{fig:tr_sigma} and Fig.~\ref{fig:tr_sigma_y},
both pump-probe configurations show characteristic features
in the same photon energy region around 15~eV.
According to the above study with the electron-hole pair distribution
in Fig.~\ref{fig:ph_map_hbn} and the two-band model analysis,
these features might also be related to the anomalous dispersion of the negative-mass
electron-hole band around the $M$ point.

Because the two-band model has spatial inversion symmetry and does not have
any even-order nonlinearity, the present model is not appropriate to investigate 
the $\omega_{pump}$-oscillation in the spectrum
of Fig.~\ref{fig:tr_sigma_y}. Therefore, an extension of the model 
would be required
to understand the microscopic mechanism of the $\omega_{pump}$-oscillation structure.

%%%%%%%%%%%%%%%%%%%%%%%%%%%%%%%%%%%%%%%%%%
\section{Summary and conclusions \label{sec:summary}}

In this work, we extended the \textit{ab-initio} pump-probe simulation of ATAS
for 2D materials in order to provide a platform to understand the microscopic origin
of the features observed in ATAS experiments.
As an example, we investigated ultrafast modification of
optical properties of monolayer \textit{h}-BN under intense IR pulses.

First, we performed the pump-probe simulations, setting the polarization of
the pump and probe pulses along the $x$-direction in Fig.~\ref{fig:current_lin}~(a),
which is perpendicular to a B-N bond.
As a result of the simulation, we obtained the transient absorption spectrum
during the laser irradiation. The resulting spectrum presented
an oscillatory behavior with a V-shaped energy
dispersion; the \textit{fishbone} structure.
The frequency of the oscillatory behavior was found to be twice the frequency of the pump pulse.
In previous studies, similar structures have been observed  in bulk materials
\cite{Lucchini916,Schlaepfer2018,PhysRevLett.117.277402}
and understood in terms of the dynamical Franz-Keldysh effect
\cite{PhysRevLett.76.4576,PhysRevB.93.045124}.

To clarify the microscopic mechanism of the observed transient absorption spectrum
of \textit{h}-BN, we studied the electronic structure of the material
with the DFT and found relevant
transitions at the $M$ and $\Gamma$ points. To elucidate the contribution
from these transitions, we further analyzed the transient absorption spectra
with a parabolic two-band model.
As a result, we identified that the fishbone structure in Fig.~\ref{fig:tr_sigma} 
consists of two major contributions. One is the dynamical Franz-Keldysh effect
around the $\Gamma$ point. The other is also the dynamical Franz-Keldysh effect
but with an anomalous band dispersion around the $M$ point.
Therefore, we clearly demonstrated that the IR-induced intraband transitions
are significant for the transient absorption even in 2D materials.
Furthermore, the discovered mechanism based on intraband motion
with the anomalous band dispersion indicates the possibility of introducing
exotic dynamical properties of the 2D material since the classical analog
of the intraband motion in the anomalous dispersion
would be a classical motion of a negative mass particle.

Next, we performed the TDDFT pump-probe simulation, setting the polarization
of the pump and probe pulses along the $y$-direction. For this direction,
the inversion symmetry of the material is broken, and the system displays
even-order nonlinear responses. Indeed, reflecting the nature of the even-order
nonlinearity, the obtained transient absorption spectrum shows
an oscillation with the frequency of the pump pulse, $\omega_{pump}$.
Therefore, it was demonstrated that the system presents a qualitatively different behavior
for different polarization directions.
However, despite of the significant differences for the different polarizations
in Fig.~\ref{fig:tr_sigma} and Fig.~\ref{fig:tr_sigma_y},
still the characteristic features emerge around the probe photon energy of 15~eV.
This fact indicates that the $\omega_{pump}$-oscillation structure in 
the transient absorption spectrum with the $y$-direction
polarization may also be related to the anomalous band dispersion at the $M$ point.
In order to understand the microscopic origin of the $\omega$-oscillation structure,
an extension of the parabolic two-band model is required.

The demonstrated \textit{ab-initio} ATAS simulation for 2D materials
is a very general approach that is applicable not only to single-layered materials
but also to multi-layered systems including heterostructures.
Here, the interplay of multi-layer and interface effects with geometrical tilting,
such as moir\'e patterns, could strongly affect the electron dynamics
and be observed in the time domain.

%%%%%%%%%%%%%%%%%%%%%%%%%%%%%%%%%%%%%%%%%%
\vspace{6pt} 

%%%%%%%%%%%%%%%%%%%%%%%%%%%%%%%%%%%%%%%%%%
%% optional
%\supplementary{The following are available online at \linksupplementary{s1}, Figure S1: title, Table S1: title, Video S1: title.}

% Only for the journal Methods and Protocols:
% If you wish to submit a video article, please do so with any other supplementary material.
% \supplementary{The following are available at \linksupplementary, Figure S1: title, Table S1: title, Video S1: title. A supporting video article is available at doi: link.}

%%%%%%%%%%%%%%%%%%%%%%%%%%%%%%%%%%%%%%%%%%
\authorcontributions{S.A.S performed the calculations. All the authors participated the interpretation
of the results and wrote the paper.}

%%%%%%%%%%%%%%%%%%%%%%%%%%%%%%%%%%%%%%%%%%
%\funding{Please add: ``This research received no external funding'' or ``This research was funded by [name of funder] grant number [xxx].'' Check carefully that the details given are accurate and use the standard spelling of funding agency names at \url{https://search.crossref.org/funding}, any errors may affect your future funding.}

%%%%%%%%%%%%%%%%%%%%%%%%%%%%%%%%%%%%%%%%%%
\acknowledgments{
We acknowledge financial support from the European Research Council 
(ERC- 2015-AdG-694097), Grupos Consolidados (IT578-13) and 
the European Union's Horizon 2020 Research and Innovation program under 
Grant Agreements no. 676580 (NOMAD). 
The Flatiron Institute is a division of the Simons Foundation.
S.A.S. acknowledges support by Alexander von Humboldt Foundation. 
}

%%%%%%%%%%%%%%%%%%%%%%%%%%%%%%%%%%%%%%%%%%
\conflictsofinterest{
The authors declare no conflict of interest.} 

%%%%%%%%%%%%%%%%%%%%%%%%%%%%%%%%%%%%%%%%%%
%% optional
\abbreviations{The following abbreviations are used in this manuscript:\\

\noindent 
\begin{tabular}{@{}ll}
ATAS & Attosecond transient absorption spectroscopy\\
2D & two dimensional\\
\textit{h}-BN & Hexagonal boron nitride\\
TMDs & Transition metal dichalcogenides\\
TDDFT & Time-dependent density functional theory\\
IR  & Infrared\\
ALDA  & Adiabatic local density approximation\\
DFT   & Density functional theory\\
FWHM  & Full width at half maximum
\end{tabular}}

\externalbibliography{yes}
\bibliography{ref.bib}

%%%%%%%%%%%%%%%%%%%%%%%%%%%%%%%%%%%%%%%%%%
%% optional
%\sampleavailability{Samples of the compounds ...... are available from the authors.}

%% for journal Sci
%\reviewreports{\\
%Reviewer 1 comments and authors’ response\\
%Reviewer 2 comments and authors’ response\\
%Reviewer 3 comments and authors’ response
%}

%%%%%%%%%%%%%%%%%%%%%%%%%%%%%%%%%%%%%%%%%%
\end{document}